# Mid-Infrared Single Photon Counting

Guilherme Temporão[1], Sébastien Tanzilli[1], Hugo Zbinden[1], Nicolas Gisin[1]

Thierry Aellen[2], Marcella Giovannini[2] and Jérome Faist[2]

[1] Group of Applied Physics, University of Geneva, 1211 Geneva 4, Switzerland

[2] Institute of Physics, University of Neuchâtel, 2000 Neuchâtel, Switzerland

We report a procedure to detect mid-infrared single photons at 4.65 µm via a two-stage scheme based on Sum Frequency Generation, using a Periodically Poled Lithium Niobate (PPLN) nonlinear crystal and a Silicon Avalanche Photodiode. An experimental investigation shows that, in addition to a high timing resolution, this technique yields a detection sensitivity of 1.24 pW with 63mW of net pump power.

Single Photon Counting (SPC) devices have been thoroughly used in the last few years in a myriad of applications, such as imaging, metrology, astronomy, spectroscopy and quantum communications. The high popularity of this detection technology is mainly due to the high sensitivity, high timing resolution and low noise it is able to achieve, especially in the case of Silicon Avalanche Photodiodes (Si APD) [1].

Nonetheless, the existing SPC technology is available for a limited wavelength range, comprising the visible and near-infrared (NIR) windows. Recently, the advent of Quantum Cascade Laser (QCL) technology has broadened the available wavelength range for free-space optical systems, especially the entire mid-infrared (MIR) window (3 µm – 20 µm) [2].

Moreover, recent studies show that adverse atmospheric effects can be mitigated by use of sources operating inside this wavelength range [3]. These results have created a demand for sensitive and fast detectors operating in MIR wavelengths.

In this Letter, we propose a detection scheme that exploits the available SPC modules for detection of MIR radiation. The procedure comprises two stages: first, the MIR photons are up-converted to the NIR via Sum Frequency Generation (SFG) and, finally, detected with a Si APD. It should be noted that the idea of using frequency conversion for detection of IR radiation is not new [4], with experiments dating from the 1960s [5] to this moment [6].

In order to characterize the performance of this approach, three figures of merit are proposed: the *overall quantum efficiency*, the *detection sensitivity* and the *timing resolution*.

The overall quantum efficiency ($\eta_{tot}$) of the detector is defined as the probability that an incoming photon will generate a counting event. It can be factorized into a product $\eta_{tot} \equiv \eta_{SFG}\eta_{opt}\eta_{det}$, where $\eta_{det}$ is the quantum efficiency of the APD, $\eta_{opt}$ comprises the losses in all optical components and $\eta_{SFG}$, called the quantum SFG efficiency, gives the probability that a MIR photon will be up-converted inside the nonlinear crystal. Assuming no phase mismatch, the Boyd-Kleinman approximation yields [7,8]:

$$\eta_{SFG} \approx \frac{32\pi^2 d_{eff}^2 P_{pump} \exp(-\alpha L) L h}{\varepsilon_0 c \lambda_{signal}^2 \lambda_{pump} n_{SFG}^2} \quad (1)$$

where $d_{eff}$ is the effective second-order electric susceptibility of the medium, $\alpha$ is the total attenuation factor for the three beams, $L$ is the crystal length, $n_{SFG}$ is the refractive index at the sum-frequency and $h$ is the Boyd-Kleinman focusing factor [8]. For a 1 cm long PPLN crystal and a pump power of 1 Watt, expression (3) gives a SFG efficiency of 0.19%, considering perfect overlap between gaussian beams [9]. This result gives an idea of the order of magnitude that can be expected for $\eta_{tot}$.

The detection sensitivity (SNR$_0$) is a measure of the minimum signal level that can be detected. It indicates the amount of power, in Watts, to obtain a signal-to-noise ratio of unity, according to the expression:

$$SNR_0 = \frac{hc}{\lambda_{signal}\eta_{tot}}\langle n_{tot}\rangle \quad (2)$$

Where $n_{tot}$ is a random variable describing the detection noise statistics and $\langle \cdot \rangle$ denotes an average value. In our case, it comprises two statistically independent sources of noise, namely, dark counts $n_{DC}$ (inherent to the Si APD) and background noise $n_{BG}$, such that $\langle n_{tot}\rangle = \langle n_{DC}\rangle + \langle n_{BG}\rangle$. Background (optical) noise consists in all counts generated by external photons other than the signal ones. In the case of a MIR detection system, it will be mostly comprised of blackbody radiation photons.

Thermal light from blackbody radiation follows a Bose-Einstein (geometric) probability distribution, with a mean number of photons per mode given by $[\exp(h\nu/kT)-1]^{-1}$ [10]. Assuming that only one spatial mode is up-converted, the background noise count rate can be approximated as:

$$\langle n_{BG}\rangle = \eta_{tot}\int_{-\infty}^{\infty}\frac{T(\nu)d\nu}{\exp(h\nu/kT)-1} \approx \frac{\eta_{tot}\Delta\nu}{\exp(h\nu_0/kT)-1} \quad (3)$$

where $T(\nu)$ is the overall normalized transfer function of the optical components, which we approximate here as a delta function centered around $\nu_0$ with bandwidth $\Delta\nu$. Note that, for high overall efficiency values, the contribution of background noise can become much greater than the dark counts of the Si APD. A careful analysis of equations (2) and (3) shows that, in this situation where total noise is dominated by background radiation, the sensitivity will no longer depend on the overall efficiency and will reach its minimum value.

Finally, the timing resolution (or jitter) $\tau$ of the SPC module indicates the uncertainty in the arrival time of a single photon. In this setup, this value is completely determined by the Si APD, which can yield values as low as 40 ps [11].

This way, the experimental part of this work consists in the measurement of the two parameters above mentioned: the overall efficiency $\eta_{tot}$ and the total noise counts $\langle n_{tot} \rangle$, which together determine the sensitivity. The schematic depicted on Figure 1 shows the configuration used in all measurements.

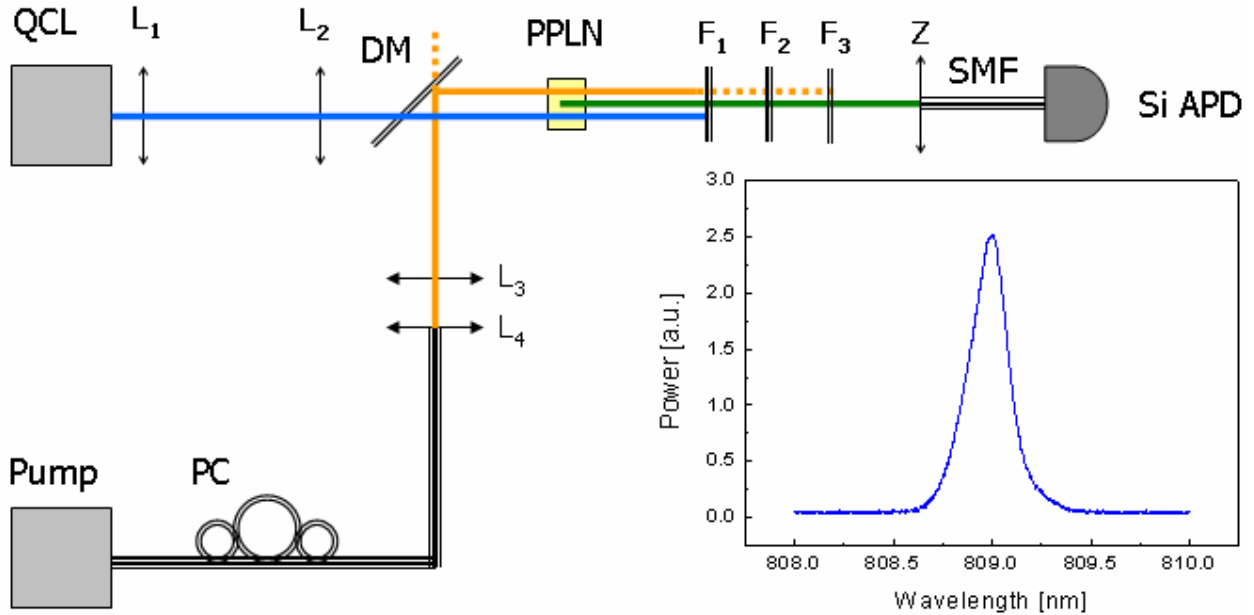

*Figure 1. Schematic of general experimental setup. See text for details. Inset: spectrum of up-converted beam immediately after the nonlinear crystal, with 0.36 nm FWHM. The resolution of the spectrometer was set to 0.1nm.*

In this setup, the QCL (from Alpes Lasers) generates 20 ns pulses of vertically-polarized MIR light at 4.65 μm, with a repetition rate of 750 kHz. Using a Dichroic Mirror (DM), this beam is combined with a pump beam of 63 mW net power, coming from a diode laser at 980 nm, inside a temperature controlled 1cm-long PPLN crystal. Phase-matching conditions have been found for two different grating periods, corresponding to crystal temperatures of 25ºC and 93ºC. A Polarization Controller (PC) is placed after the pump laser to ensure vertical polarization.

In order to remove the pump beam after the crystal, a 10nm wide band-pass filter at 810 nm (F1) was used in conjunction with a low-pass filter at 930 nm (F2). The PPLN phase-matching bandwidth of 1.47THz is limited to exactly fit the spectral width of the up-converted beam (see inset in Fig. 1) by a 0.35 nm wide (160 GHz) interference filter at 810 nm (F3) for background noise reduction. The up-converted signal is focalized inside a 830nm Single Mode Fiber (SMF), with the help of a lens (Z), and is detected by an ungated EG&G Active Quenching Si APD which was kept isolated from ambient light. The positions and focal lengths of all lenses (L1-L4) were selected such that the beam parameters for the signal and pump beams are as close as possible to the conditions of maximum overlap [7,8].

In the first measurement, the QCL was attenuated to an average power of a few nanowatts. The signal photon rate $n_{QCL}$, the total detection rate $n_{det}$ and the detection efficiencies are summarized in Table 1.

Table 1. Detection Efficiency Results

| $n_{QCL}$ [kHz] | $n_{det}$ [kHz] | $\eta_{tot} = \dfrac{n_{det}}{n_{QCL}}$ | $\eta_{opt}$ | $\eta_{det}$ | $\eta_{SFG} = \dfrac{\eta_{tot}}{\eta_{opt}\eta_{det}}$ |
|---|---|---|---|---|---|
| $3.6 \cdot 10^7$ | $125 \pm 1$ | $3.6 \times 10^{-6}$ | 13.7% | 52% | $5.06 \times 10^{-5}$ |

The up-conversion efficiency $\eta_{SFG}$ was found to be $5.06 \times 10^{-5}$, which corresponds to $8.03 \times 10^{-4}$ per Watt of pump power. This is off by a factor 2.4 with respect to the theoretical value. This difference can be assigned to the non-negligible absorption of 4.65 μm radiation by Lithium Niobate (measured to be around 40%), which deviates the optimal focusing conditions from the ideal situation [7].

The second goal of the experiment was to characterize the noise rates. Whereas the dark counts are measured with all sources switched off, the background noise is measured by switching off the QCL but keeping the pump beam on. This measurement has been performed for the two phase-matching crystal temperatures, and the values are presented in Table 2.

Table 2. Total Noise Counts [Hz]

|  | 25°C | 93°C |
|---|---|---|
| $\langle n_{DC} \rangle$ | 55.0 | 55.0 |
| $\langle n_{tot} \rangle = \langle n_{DC} \rangle + \langle n_{BG} \rangle$ | 87.8 | 133.1 |

Since the crystal is not completely transparent at 4.65μm, a higher temperature yields a higher number of blackbody photons that are created inside the crystal in the same spatial mode. These values also comprise residual pump photons that manage to bypass the filters.

Now that the system has been characterized, the QCL can be calibrated to generate only one photon per pulse. Considering a pulse rate of 750 kHz and the overall efficiency value in Table 1, this corresponds to a detection rate of 2.7 Hz. In order to make the noise counts negligible, the QCL pulses were reduced to 1 ns and a Time to Amplitude Converter (TAC) was used to make coincidence counts between detection events in the Si APD and electrical pulses in the QCL. The result can be seen in Figure 2: after several minutes, one can reconstruct the non-attenuated pulse shape of up-converted light. It's important to realize that, even though only 3 photons out of 750 thousand are detected, one has the precise information about which one-photon pulses have been detected. This is a feature unique to photon counters, and crucial for quantum communications applications.

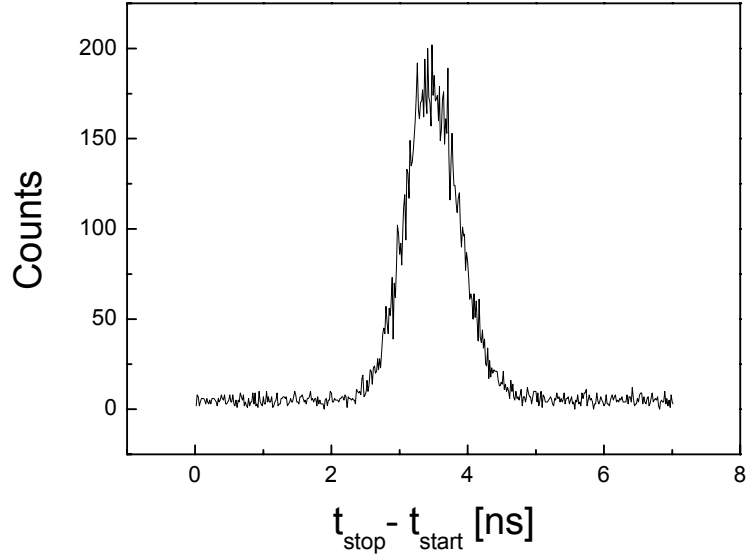

*Figure 2. Histogram of photon arrival times ($t_{start}$) with respect to QCL electric pulses ($t_{stop}$) after several minutes.*

Finally, one can now calculate the $SNR_0$ using expression (2), which yields 2.82 pW at 93°C and 1.24 pW at 25°C. Table 3, below, shows a comparison for the sensitivity and the time response between our results and two commercially available MIR detectors at this wavelength: a room-temperature detector (Vigo System PVI-5) and a state-of-the-art liquid nitrogen cooled detector (Fermionics PV-650).

| | τ [ns] | SNR$_0$ [pW] |
|---|---|---|
| Vigo System PVI-5 | 15 | $1.63 \times 10^6$ |
| Fermionics PV-650 | 20 | 223 |
| PPLN @ 25°C + EG&G AQR-15FC | 0.3* | 1.24 |

Table 3. Comparison with commercially Available MIR detectors at 4.65 μm

* timing resolution

In conclusion, it has been shown that the frequency up-conversion approach for photon counting in the MIR yields a single-mode, fast and sensitive detector, despite the low quantum efficiency. If there is a need for higher efficiency values, this can be done by either increasing the pump power or the transmission of the optical components, which will also decrease the sensitivity. Moreover, using a coincidence count technique, we have been able to demonstrate the detection of MIR light pulses containing only one photon.

We would like to thank NCCR Quantum Photonics, CAPES and EXFO Inc. for the financial support.

## References

1. S. Cova *et al*, Rev. Sci. Instrum. **60**, p. 1104 (1989)

2. J. Faist *et al*, Appl. Phys. Lett. **68**, p. 3680 (1996).

3. H. Manor *et al*, Appl. Optics **42**, p. 4285 (2003)

4. N. Bloembergen, Phys. Rev. Lett. **2**, p. 84 (1959)

5. J. Midwinter and J. Warner, J. Appl. Phys. **38**, p. 519 (1967)

6. K. Karstad *et al*, Optics and Lasers in Engineering **43**, p. 537 (2005)

7. G. D. Boyd and D. A. Kleinman, J. Appl. Phys. **39**, p. 3597 (1968)

8. W. Risk, T. Gosnell and A. Nurmikko, Compact Blue-Green Lasers, Cambridge, 2003

9.  We are assuming here $d_{eff}$ = 16 pm/V for PPLN

10. B. Saleh and M. Teich, Fundamentals of Photonics, Wiley, 1991

11. See, for example, www.idquantique.com